Encyclopedia of Astrophysics 1st Edition

# Chapter 10031. Surfaces and Interiors

**Authors**

Lena Noack, Freie Universität Berlin, Department of Earth Sciences, Malteserstr. 74-100, 12249 Berlin, Germany, lena.noack@fu-berlin.de, Corresponding author

Caroline Dorn, ETH Zurich, Department of Physics, Wolfgang-Pauli-Str. 27, 8093 Zurich, Switzerland, dornc@phys.ethz.ch

Philipp Baumeister, Freie Universität Berlin, Department of Earth Sciences, Malteserstr. 74-100, 12249 Berlin, Germany, philipp.baumeister@fu-berlin.de

**Abstract**

In the last 15 years, since the discovery of the first low-mass planets beyond the solar system, there has been tremendous progress in understanding the diversity of (super-)Earth and sub-Neptune exoplanets. Especially the influence of the planetary interior on the surface evolution (including the atmosphere) of exoplanets has been studied in detail. The first studies focused on the characterization of planets, including their potential interior structure, using as key observables only mass and radius. Meanwhile, a new field of geosciences of exoplanets has emerged, linking the planet to its stellar environment, and by coupling interior chemistry and dynamics to surface regimes and atmospheric compositions.

The new era of atmospheric characterization by JWST as well as the ELT will allow testing of these theoretical predictions of atmospheric diversity based on interior structure, evolution, and outgassing models.

**Keywords**

- Exoplanet structure
- Planetary interior
- Chemical composition
- Super Earths
- Star-planet interactions
- Mantle
- Exoplanet atmospheric evolution

**Chapter Outline**

1. Introduction
2. Composition and interior structure of exoplanets
    2.1. Star-planet connection
    2.2. Equations of state
    2.3. Interior structure
3. Feedback between interior and surface
    3.1. Interior as global heat engine
    3.2. Interior as driver of planetary dynamics
    3.3. Interior as source of crust and atmosphere
4. Conclusion

## Glossary and Nomenclature

| | |
|---|---|
| CMB | Core-mantle boundary for a differentiated body. |
| CME | Coronal Mass Ejections that are large eruptions of a star, ejecting plasma from the star's corona. |
| ELT | *Extremely Large Telescope*, European Southern Observatory (ESO), Chile, planned to be operational from the late 2020's on. |
| EoS | Equation of State to derive thermodynamic properties for a given material. |
| Equilibrium temperature | Theoretical temperature that a planet would have in radiative equilibrium, i.e. the temperature at which the incident solar radiation would be in equilibrium with outgoing radiation, without considering an atmosphere. Due to greenhouse gases, the surface temperature can be very different from the equilibrium temperature. |
| JWST | *James Webb Space Telescope*, NASA (National Aeronautics and Space Administration, US), operational in Earth's orbit since 2022. |
| Love numbers | Love numbers ($k, h, l$) are non-dimensional parameters describing the rigidity and deformation of a planetary body. |
| M dwarf | Low-mass star. Main-sequence stars are classified in 7 classes of increasing temperature starting with M stars, followed by K, G, F, A, B, and O stars. Our Sun is a G-type star. |
| PLATO | *PLAnetary Transits and Oscillations of stars*, ESA (European Space Agency), space mission planned to be launched into Earth's orbit in 2026. |
| Redox state | Oxidation state of a material, varying from *red*ucing to *ox*idizing conditions. A reducing environment is characterized by little or no free oxygen. |
| Refractory | Refractory elements have a very high condensation temperature. Terrestrial planets are mostly made of refractory elements (we here define refractory elements beginning with the condensation temperatures of Mg, Si and Fe). For rocky planets with very hot surfaces, silicate atmospheres may form that are made of refractory elements. |
| Sub-Neptune | One of the most common classes of exoplanets discovered to date with radii below Neptune but larger than a primarily rocky body, implying a volatile-rich envelope. |
| Super-Earth | The other most common class of exoplanets discovered to date with radii and masses above Earth's values but with densities that would imply a rocky composition. |
| Super-Mercury | Sub-class of super-Earth planets with a very high densities, typically explained by a high metal enrichment. |
| TRAPPIST | *TRAnsiting Planets and PlanetesImals Small Telescope*, pair of ground-based telescopes situated in Chile (TRAPPIST-South) and Morocco (TRAPPIST-North). Discovered the TRAPPIST-1 planetary system with 7 planets of approximately Earth's size. |
| Volatile | Volatile elements condense/solidify at low temperatures. Planets in the outer solar system are made of both volatile and refractory elements. Atmospheres of planets are typically made of highly volatile elements (H, C, N, O, noble gases), termed volatiles. |

**Learning Objectives**

- Exoplanets should show a wide diversity regarding their interior composition in relationship to the observed stellar compositional diversity as well as their accretion history.
- The planetary interior is expected to strongly influence the surface evolution (including the atmosphere) of exoplanets - both in terms of surface regime (e.g. plate tectonics) as well as atmospheric evolution.
- To understand the diversity of exoplanetary interiors, models no longer rely on mass and radius measurements alone, but also include information on the stellar environment and potential thermal evolution of the exoplanet.
- As the interior is linked to the atmosphere, atmospheric characterization by JWST as well as the ELT will allow to test theoretical predictions of atmospheric diversity based on interior structure, evolution, and outgassing models.

## 1. Introduction

In the past thirty years, the number of known planets changed dramatically, from nine (eight) planets within our solar system to more than 5000 unambiguously detected worlds around other thousands of stars. What have we learned in these past decades, since the first confirmed discoveries of exoplanets around solar-like stars in the early 1990's (Mayor and Queloz, 1995)?

The large sample size allows us to make statistical arguments and to put our own planetary system into a much larger context. The first discoveries of exoplanets were strongly biased towards close-in and massive planets (so-called "Hot Jupiters") due to their improved detectability, but in recent years, mostly thanks to the Kepler survey, we have found many planetary systems that resemble our own, with rocky planets close-in and gas giants and Neptunes further out. At the same time, we have also discovered many systems that lack either small and potentially rocky planets (e.g. due to migration of gas giants) or in turn only consist of roughly Earth-size bodies (especially around M dwarfs, including for example the famous TRAPPIST-1 system, Gillon et al., 2017). Our own system therefore seems to be neither the rule nor the exception, but only one example of system architectures, that can be immensely diverse depending on the stellar properties. On the other hand, our current observations are still strongly limited and biased, and future long-term observational strategies from ground as well as new space telescopes such as PLATO will extend our current statistical view on exoplanets.

One important finding of the past two decades, however, was that planets are not restricted to distinct planetary classes observed in our solar system (with a large gap in size/mass between rocky planets of maximum Earth' size in the inner solar system and the ice or gas giants in the outer solar system), but that the parameter space in between is filled with a large number of super-Earth and sub-Neptune planets, whose exact nature we still do not fully understand. We still do see a clustering of detected planets around low-mass (up to a few Earth masses) and high-mass planets, with a gap in-between in radius (Fulton et al., 2017), which, however, is much smaller than observed in the solar system and whose extent also strongly depends on the stellar type. By now an observational bias can be excluded and theories on planet formation and early evolution (especially atmosphere losses for sub-Neptune planets) have been developed that can explain the observed exoplanet populations (Owen and Schlichting, 2024).

In addition to statistical arguments, selected multi-planetary systems also allow for a more elaborate characterization of planets, including for example their potential compositional variation within one system depending on their orbital distances (e.g. Acuña et al., 2022), including the previously mentioned TRAPPIST-1 system, but also TOI-178 (a 7 Gyr old system) or Kepler-80 (a 2 Gyr old system). Variations in density can then be attributed to differences in composition, which specifically would hint at either different accretion histories (especially in the case of super-Mercuries showing larger-than expected core-mass fractions, see Section 2.1), variations in ice/water fractions, or the existence of different types of atmospheres including extended, low-density atmospheres (such as primordial $H_2$-He atmospheres).

But data on exoplanets is not limited to mass and radius, or age of the system. Next to orbital information that can inform us about expected effective surface temperature (i.e. in the absence of greenhouse gases) and tidal forces leading to additional energy dissipation in the planet's interior, we can also obtain a first order estimate on a planet's composition from the stellar metallicity and chemical abundances in the stellar spectrum. Emission spectroscopy (either via direct imaging or as a complement to the stellar signal during a secondary eclipse) can give us valuable information about the surface or atmosphere of a body, which in the era of the JWST is allowing us a large step forward towards the characterization of planets beyond our solar system.

The wealth of information that we can now gain about an exoplanet are (in an ideal scenario) for example comparable to the data obtained for the moons of Jupiter before they were visited by the first spacecraft *Pioneer 10*: remote determination of orbital information as well as albedo measurements and spectroscopic data allowed for a first characterization of Io as a rocky moon including a sulfur-rich surface (Lee et al., 1972) in contrast to the other icy moons Europa, Ganymede and Callisto, where the brightness of these bodies already allowed for a first indication of the age of the ice (with Europa having a fresh water-ice crust and Callisto having an old crust composed of a mixture of ice and dust). While the age of spaceflight did allow for astonishing discoveries (including the previously underestimated strength of tidal heating inside of Io, as well as the discovery of subsurface water on Europa, Ganymede and maybe even Callisto), the first-order remote characterization matched our current knowledge of these moons surprisingly well, which is promising with respect to the interpretation of current and future exoplanet observations.

These interpretations are aided by modeling approaches to better understand interior and surface processes or exoplanets, where both modeling and experimental advances in the past 20 years since the first discovery of low-mass rocky exoplanets (such as CoRot-7b and Kepler-10b) allowed characterization of exoplanets to grow out of its childhood into a mature research field. These advances were only possible due to interdisciplinary collaborations connecting the dots between the stellar environment of planets with state-of-the-art experimental data and looking-beyond-the-boundaries of solar system knowledge to begin to grasp the diversity of exoplanetary interiors that can be out there – but learning from previous experience it is clear, that many discoveries still await us!

## 2. Composition and interior structure of exoplanets
### 2.1. Star-planet connection

Numerical models of planet formation that examine equilibrium condensation sequences propose that planets inherit some chemical make-up of their host stars. For example, Thiabaud et al. (2015) illustrate that most planets exhibit a bulk refractory composition similar to their host star, specifically for the rock-building elements of Fe, Si, and Mg. They have high condensation temperatures (>1000 K), such that refractory species (e.g., oxide species) condense close to the host star in a protoplanetary disk. In consequence, planets tend to replicate the refractory element ratios of the protoplanetary disk. Elements like Mg, Si, and Fe are observable in stellar photospheres, and their ratios are useful constraints for planet interior modeling. When these constraints are applied to interior models, they suggest that the mantle typically forms the largest layer, as opposed to the iron core, in most rocky planets. However, there is still debate about the extent to which stellar abundance proxies can accurately inform planetary rock compositions (Dorn et al., 2015, Schulze et al., 2021, Plotnykov and Valencia, 2020). Adibekyan et al. (2021) found a correlation between the compositions of rocky planets and their host stars, indicating that the relationship for Fe/Mg is not exactly 1:1, and that planets can be richer in iron than would be expected from their host stars. However, for a final conclusion there is a need for robust and comparable stellar abundance estimates. Spectra of white dwarfs that were polluted by recently accreted materials (broken-up parts of asteroids, in other words planetary building materials, or even remnants of planetary bodies themselves) can give additional chemical constraints on the compositional variety of exoplanets.

While the main rock-forming elements Mg, Si, and Fe show similar condensation temperatures (Lodders, 2003), other refractory as well as volatile elements condense from the nebula at very different temperatures, leading to various different condensation or ice lines and therefore compositional variations in the planetary building blocks depending on the distance to the host star (devolatilization trend, Wang et al, 2019). However, especially for volatile species, a direct link between the condensation ice lines and later planetary compositions is not straight-forward due to several secondary processes including pebble migration, disk evolution and instabilities, devolatilization during planetesimal accretion, and outgassing during early melting events in proto-planetary bodies. Observational constraints on the link between stellar and planetary composition therefore focuses on close-in planets, including extreme cases such as super-Mercuries as well as strongly heated magma ocean planets.

Adibekyan et al. (2021) showed, for example, that super-Earths and super-Mercuries might be distinct populations, suggesting that the latter may not be formed by giant impacts as often proposed. Apparently, the collisional history of planet formation does not explain the observed diversity in planet density. While giant impacts are one possible component to form super-Mercuries, there are other possibilities which have been explored on how to form super-Mercuries. These include, for example, nucleation and growth processes of iron pebbles (Johansen & Dorn, 2022). Mercury in the solar system shows anomalous characteristics. Its origin is still debated and so far, no single process (e.g., condensation sequence, giant impact accretion processes) has been identified to explain all the observed features (e.g., lack of FeO, reduced oxidation state of crust and mantle, moderately volatile elements present on surface). For exoplanets, as we are probing predominantly close-in planets, compositionally extreme worlds may be found that form in high-temperature regimes. Using the condensation sequence of proto-planetary gas disks (Dorn et al., 2019) have identified a potential class of exoplanets that forms from high-temperature condensates (iron-poor and rich in Ca- and Al-oxides) and whose bulk densities are lower compared to Earth-like compositions. Their existence may be verified by atmospheric characterization. Similarly, Plotnykov and Valencia (2020) have shown that the statistical range of possible abundances of rocky planets scatters wider than host star abundances, including potentially Fe-Si-depleted rocky planets. On the other hand, close-in low-density planets may also be explained by large fractions of melt being less dense than solid rock.

Most super-Earths are hot worlds which reside within the moist greenhouse radiation limit (> 400 K equilibrium temperature). This implies that any atmosphere may increase surface temperatures drastically to allow for molten silicates, i.e., a magma ocean. For a hand-full of super-Earths this is even true without any atmosphere where equilibrium temperatures are above ~1800 K. Hence, the majority of the observed super-Earth population is dominated by long-lived magma oceans. The boundary between magma ocean and atmosphere is compositionally coupled, chemically reactive, and thermally active (e.g. Kite et al., 2020).

Future detections of rocky exoplanets on longer orbits around F- and G-type stars with PLATO will allow us to obtain an improved view on potential compositional links between star and planet at different condensation regimes and deviations from the current predictions, since our current view is heavily biased towards M-dwarf systems. Comparative planetology in multi-planet systems will allow for the constraint or refutation of density trends within planetary systems, that can then be linked with theories on planet formation and migration.

## 2.2. Equations of state

Building upon the studies of the interior structure of planets in the Solar System, models aimed at characterizing the interior of low-mass exoplanets generally assume that a planet consists of layers with physical and chemical properties. Rocky (terrestrial) planets are dominated by silicates and iron-rich cores. Planets with lower densities likely contain significant amounts of volatile elements, such as hydrogen-rich atmospheres or water.

To first order, planets are spherically symmetric and in hydrostatic equilibrium. Under this assumption, the interior structure of a planet can be characterized by a set of 1D fundamental structural equations which link mass $m$, radius $r$, density $\rho$, pressure $P$, and temperature $T$ inside the planet, depending on the gravitational constant $G$ and composition $c$:

(1) $\frac{dm(r)}{dr} = 4\pi r^2 \rho(r)$

(2) $\frac{dP(r)}{dr} = -\frac{Gm(r)\rho(r)}{r^2}$

(3) $P(r) = f(\rho, T, c, \dots)$

Central to the modeling of the interior structure of planets are the Equations of State (EoS, equation 3), that describe the relationship between thermodynamic parameters such as density with pressure and temperature for a given material via a function $f$. In the context of rocky planets, equations of state typically stem from thermodynamic theoretical models, which are fitted to experimental data, for example from high-pressure diamond anvil cells and laser shock compression experiments. Numerical models, e.g. ab-initio calculations are employed for extreme conditions unachievable in the laboratory.

EoS are usually valid only over the specific range of the experimental data, which is often exceeded in planetary conditions. This necessitates the extrapolation of the EoS to extreme conditions, where predicted material properties may not be reliable anymore. This leads to typically small, but non-negligible uncertainties in the characterization of interior properties. Hakim et al. (2018) demonstrate that in the TPa range, the extrapolation of Fe EoS can lead to differences in inferred iron density of up to 20%, which results in modeled radius uncertainties of several percent. However, overall, these effects are generally minor compared to the uncertainties arising from compositional unknowns. Nevertheless, special care should be taken to select EoS appropriate to the pressure and temperature conditions in the planet.

Aguichine et al. (2021) find similar results for water-rich planets. They find that extrapolating water EoS outside their defined range can lead to an overestimation of the planetary radius by up to 10%. These conditions occur for water mass fractions above 5%, which necessitates the use of a water EoS which holds up to a few TPa and several thousand Kelvin.

One influential compositional unknown is the amount of light elements in the iron core. It is well known that the density of Earth's core is too low for it to be pure iron. This density deficit can be explained by the presence of lighter elements, such as hydrogen, sulfur, or carbon, but the exact nature of the density deficit is still unknown even with the wealth of information available on Earth. It is likely that the presence of lighter elements in exoplanet cores is the norm. This introduces a significant compositional degeneracy in the interior structures.

Hakim et al. (2018) show that the density deficit arising from lighter elements can significantly impact the interpretation of the retrieved interior structure. A 20% reduction in core density decreases the modeled mass of a planet by 10-30%, depending on core size.

The appropriate choice of EoS is particularly important for volatile phases, such as water layers or atmospheres. Depending on the irradiation received from the host star, massive water-rich planets may display supercritical water layers surrounded by thick steam atmospheres instead of high-pressure ice phases (Aguichine et al. 2021), which offers an additional explanation for the radius gap.

Given the uncertainties regarding mass, radius and composition of a planet, the temperature profiles play only a minor role in silicate and iron layers, but can be significant for volatile-rich layers such as water layers (Thomas and Madhusudhan, 2016) and the atmosphere (Turbet et al., 2020).

### 2.3. Interior structure

Super-Earths and sub-Neptunes are the most common observed exoplanets. They show significant variability in their mass and radius. Super-Earths, with radii smaller than about 1.5 Earth radii, are thought to have rocky compositions, but they may also contain low amounts of water or other volatiles. Sub-Neptunes, with radii between approximately 2 and 3 Earth radii, likely have thick hydrogen/helium envelopes and may also host extensive water layers mixed within their envelopes (e.g., Kite et al., 2020, Schlichting and Young, 2022).

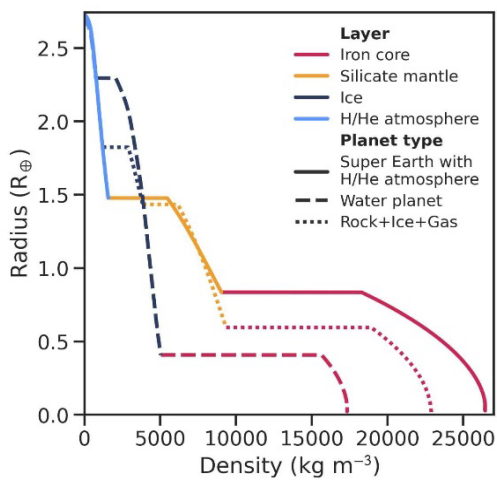

Fig. 1: Three qualitatively different interior structures, which all fit the observed parameters of the well-studied exoplanet GJ 1214 b. The planet's interior could e.g. be explained as an iron- and silicate-rich super-Earth with an extended H/He envelope (solid line), as a volatile-rich water (dashed line) world, or by some mixture of the two (dotted line).

There is an inherent degeneracy in interior modeling, meaning that a given set of observables (mass and radius) can correspond to multiple different interior compositions and structures, as illustrated in Figure 1. To reduce this degeneracy, additional constraints and data are necessary that come from a large variety of sources, including first principle considerations on the general composition of planets, lab experiments, ab-initio calculations, stellar properties, planet system architecture and tidal dissipation considerations, or planet formation. Among the most important constraints on the interiors of super-Earths and sub-Neptunes is the bimodal distribution of planet sizes (Fulton et al., 2017). Between around 1.5-2 Earth radii, there is a clear scarcity of planets, which has been interpreted to be due to evolution processes: sub-Neptunes have thick H/He atmospheres, while super-Earths have lost them and represent the stripped interiors. Both core-powered mass loss and evaporative loss shape the super-Earths population, while the radius valley itself is carved by photoevaporative loss (Owen & Schlichting, 2023). The interiors of super-Earths are not necessarily just the H/He-stripped counterparts of sub-Neptunes. Super-Earths and sub-Neptunes may differ also in terms of their water budgets by formation (Venturini et al., 2020). In fact, formation models predict that water-rich worlds are efficiently migrating inwards to locations where we observe them today. That said, the radius valley as the most important constraint on the planet interiors is (not yet) imposing strong enough constraints to reduce the inherent degeneracy.

Recent advancements in exoplanet science have emphasized the importance of considering complex interactions between a planet's interior and its atmosphere. Many super-Earths and sub-Neptunes are likely to have magma oceans even at their evolved stages. In addition, even temperate rock-dominated planets start their evolution hot, mainly because of the release of gravitational potential energy from accretion, radiogenic heating and differentiation. Several processes act in magma oceans with implications on the observable atmosphere of super-Earths. Melting and solidification, outgassing, redox-reactions, core differentiation and the loss of atmospheres to space are governing processes in this early stage of a planet's evolution.

To address these challenges, interior models are developed that account for chemical and compositional coupling between atmosphere and deeper interior (e.g. Kite et al., 2020; Schlichting and Young, 2022; Baumeister et al., 2023). Such coupling is crucial for accurately estimating volatile inventories, which inform us about a planet's formation environment, evolutionary history, redox state, and potential habitability. For instance, the presence of water and other volatiles can significantly affect melting temperatures of silicates, interior structure and atmospheric structure.

As mentioned above, one of the most intriguing systems for studying exoplanet interiors is the TRAPPIST-1 system, which contains seven Earth-sized planets orbiting an ultracool dwarf star. The TRAPPIST-1 planets offer a unique opportunity to study a diverse set of planetary interiors within the same system. Observations suggest that these planets have densities lower than Earth, indicating the presence of volatiles or less dense refractory materials. The exact compositions of these planets remain uncertain, but models suggest they could have a range of water content, from dry, rocky planets to those with substantial water envelopes (Agol et al., 2021). For instance, TRAPPIST-1b, the inner-most planet of the system, is thought to have a rocky composition with no clear evidence of an atmosphere. TRAPPIST-1c may have a thin secondary outgassed atmosphere but otherwise does not seem to contain many volatile elements (Zieba et al., 2023). TRAPPIST-1f and TRAPPIST-1g might have thicker atmospheres and higher water contents. The varying compositions among the TRAPPIST-1 planets illustrate the diversity of interior structures within a planetary system.

### 3. Feedback between interior and surface

One of the main factors influencing the surface conditions of a planet (including the atmosphere) is the interior of the planet. At least for low-mass planets, the long-term evolution of the atmosphere is driven by volcanic outgassing as well as (at least in the case of Earth) recycling of volatiles back into the interior via plate tectonics. The composition of the interior as well as remelting processes of crustal material (for example via subduction) define the composition of the surface material - which is what we may, at least for some exoplanets, observe remotely. But also distinct features at the surface such as a dichotomy between lowlands and highlands as on Mars, crustal compositional variations as on Venus, or the possibility to have a partial coverage of the surface by water depending on the topography and the global carbon cycle as on Earth, are all influenced by interior processes. For a correct interpretation of observational features in the emission or transmission spectrum, or in the phase curve of an exoplanet, a good understanding of different interior processes is inevitable.

## 3.1. Interior as global heat engine

The main planetary processes that are driven by the interior include, but are not limited to, mantle convection, surface movement (for example via plate tectonics or convective mobilization), surface lava flows, volcanic outgassing, recycling of volatiles, and generation or maintenance of a magnetic field. All of these processes are directly linked to the heat sources available in the interior of a planet. We distinguish between several different sources of internal energy:

### 1) Accretional energy

This term describes the energy increase of the planet by transferring kinetic energy of accreted material to the planet. The energy is often described with the following formula, which describes the gravitational energy that is released when bringing a particle $i$ to a planet of mass $M$ and radius $R$,

$$(4) \quad E_i = C_{loss} G \frac{M \mu_i}{R}$$

where $\mu_i$ is the mass of the particle, $G$ is the universal gravitational constant, and $C_{loss}$ is a constant describing energy loss processes.

### 2) Gravitational energy due to core formation

When the heavy iron separates from the molten silicate rock to combine into iron droplets that sink towards the core due to gravitational forces, energy is released and contributes to both heating the mantle as well as the core. Following Foley et al. (2020), the release of gravitational potential energy release due to core formation alone was sufficient to increase the temperature at the core-mantle boundary of Earth by 4000 K.

### 3) Latent heat release and gravitational energy upon core crystallization

For an Earth-size or super-Earth-size body, the metal core is expected to crystallize from the interior out, leading to an inner solid core of almost pure metals and a liquid outer core enriched in lighter elements (such as sulfur). The phase transition from liquid to solid releases latent heat, which can be approximated as

$$(5) \quad E_L = L \frac{dm}{dt}$$

where $L$ is the latent heat per mass and $m$ is the increasing mass of the inner core.

Due to the release of the lighter elements from the freezing inner core into the outer liquid core, additional gravitational energy $E_G$ is released into the core and depends on the density difference between pure and enriched metals as well as the radius of inner and outer core.

### 4) Radiogenic heating

One of the main heat sources inside rocky planets after the accretion is heating by release of energy during radioactive decay, especially of the isotopes $Al^{26}, Fe^{60}, U^{235}, U^{238}, Th^{232}$ and $K^{40}$. An overview of radiogenic heat generation for each isotope as well as compositional variation between known bodies (i.e. in the solar system) is summarized in Foley et al. (2020). The first two isotopes are very short-lived and become extinct during only a few Myrs, but they may have a strong influence on potential melting processes inside planetesimals. The remaining heat sources are long-lived isotopes with half-life times between several hundreds of Myrs and more than ten Gyrs. As a result, the total radiogenic heating for an Earth-like mantle decreases by a factor of about 5 over 4.5 Gyr of evolution, see Figure 2, with heating initially being dominated by the U and K radiogenic isotopes, and long-term heating being attributed to the more stable Th isotopes due to their slow decay time.

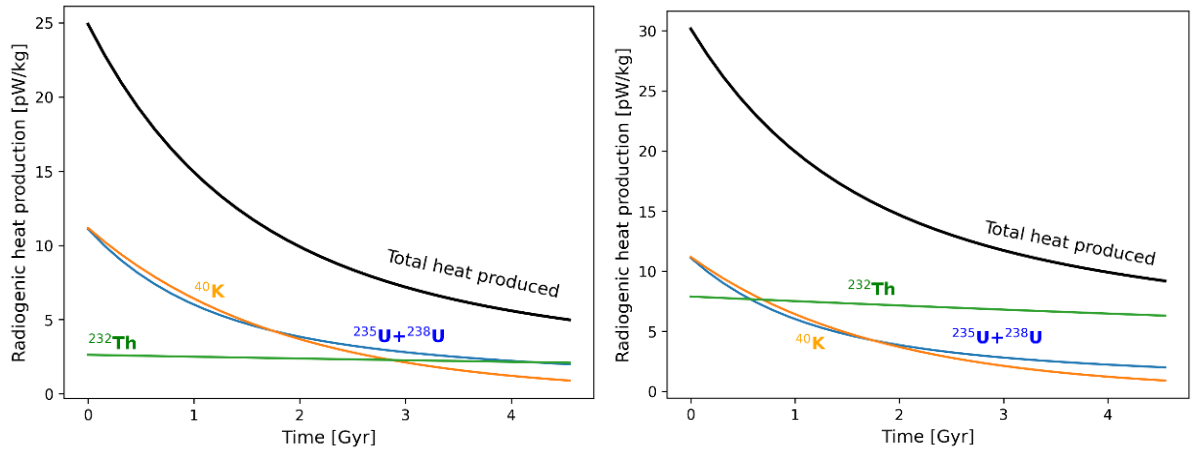

*Fig. 2: Radiogenic decay over time calculated for the main radiogenic heat sources of Earth (left) and for a model planet where the Thorium isotopes are three times as high as for Earth (right). Depending on the chemical abundances of radioactive isotopes, which can vary strongly for planets around other stars, decay of heat production may be stronger (if $^{40}K$ is the dominate radiogenic isotope) or weaker than observed for Earth (for $^{232}Th$-dominated radiogenic heating).*

**5) Tidal heating**

For close-in exoplanets or exomoons orbiting gas giants (similar to the close-in moons of Jupiter), tidal heating due to the interplay between gravitational pulls of the other bodies in the system can lead to strong tidal deformations of the body and frictional heating in the interior, leading potentially to extremely high interior temperatures and even subsurface magma oceans (as suggested for Io or the innermost TRAPPIST-1 planets, for example). The total dissipated power of tidal heating for a body in synchronous rotation can be calculated from

(6) $\dot{E} = \frac{|k_2|}{Q}\frac{(\varpi R)^5}{G}\left(\frac{21}{2}e^2 + \frac{3}{2}sin^2 I\right)$

where $|k_2|$ is the absolute value of the Love number $k_2$, $Q$ is the dissipation factor, $\varpi$ is the orbital period, $G$ the gravitational constant, $e$ the eccentricity, and $I$ is the obliquity. The dissipation factor $Q$ is a few hundred for Earth, about 100 for Mars, and a few tens for the Moon. For Io, on the other hand, $Q$ may reach values of about $10^6$. This comparison shows that there is no linear relationship between $Q$ and body size, since the interior state influences the value of $Q$ as well. Figure 3 shows an estimate of the variation in tidal energy released inside of the TRAPPIST-1 planets depending on different assumptions on mantle viscosity (for a fixed mantle shear modulus), leading to a large range of dissipation factors and tidal heating values.

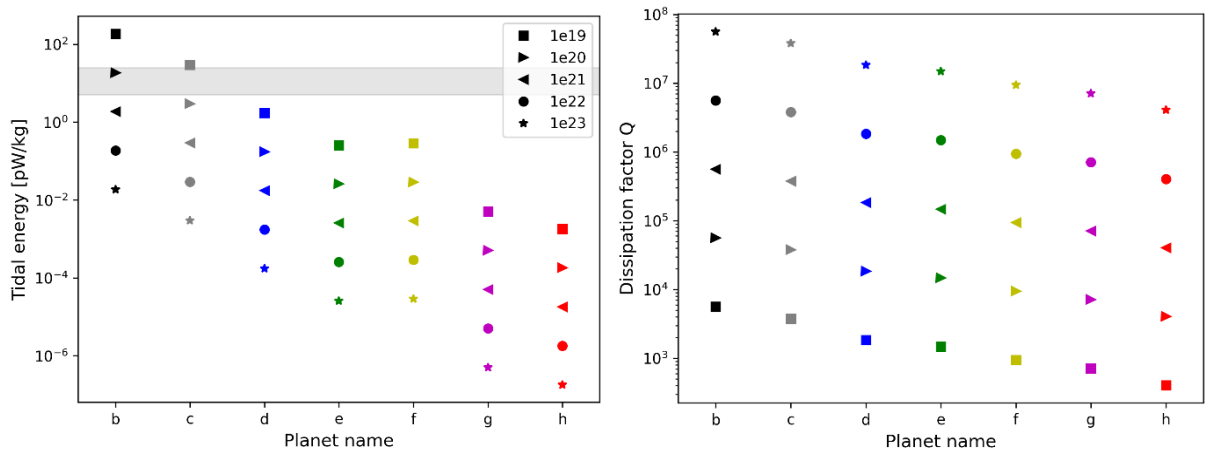

Fig. 3: Tidal heating estimates (left) and dissipation factors (right) for TRAPPIST-1 planets for different assumed mantle viscosities at a constant shear modulus of 100 GPa. The gray bar shows the range of radiogenic heating in Earth's mantle since accretion until today.

**6) Induction heating**

For close-in planets around stars with strong magnetic fields, near-surface rocks can be heated by magnetic induction heating depending on the exact orbital configuration of the planet including eccentricity and obliquity, magnetic field axis in alignment with stellar rotational axis, and orbital inclination. Electrically conductive materials in the rocky crust and lithosphere of a planet (mainly referring to hydrated crust and iron-rich minerals) can then be heated by induction when exposed to variations in the stellar magnetic field (Kislyakova et al., 2017) - similar to the principle of induction ovens on Earth which are used to melt metals.

**7) Irradiation**

Depending on the distance to the star, a non-negligible heat source may be the irradiation of the surface of the planet from its star. For exoplanets, a first indication of the resulting surface temperature is the effective temperature depending on the solar flux and albedo of the planet. A thick greenhouse atmosphere can strongly amplify the heating effect at the surface. For close-in planets, the surface temperature may thus exceed even the melting temperature of rocks, which would lead to a hemispherical magma ocean (since close-in planets would be assumed to be tidally locked, hence with a fixed hemisphere exposed to the stellar irradiation). If no atmosphere persists that transports the heat from the day-side of the planet to the night-side, the hot day-side would reduce efficient cooling from the interior of that hemisphere. Considerable temperature variations would therefore also be expected in the deep-interior, as interior mantle convection may not be fast enough to homogenize interior temperatures from day- to night-side of the planet.

All heat sources are summarized again in Figure 4.

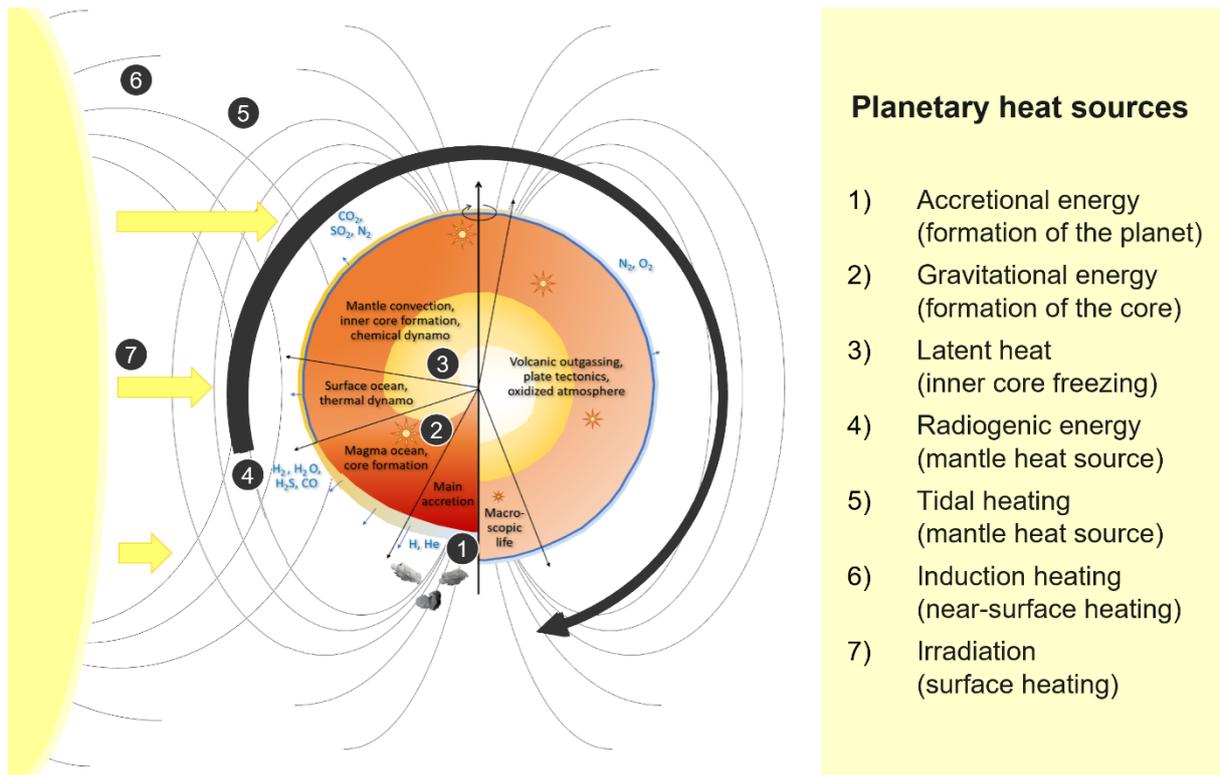

*Fig. 4: Summary of the possible heat sources of interior and surface from accretion to long-term evolution of rocky planets at the example of Earth depicted over time. Decreasing radiogenic heating in the mantle is indicated by the thick round arrow. Increasing luminosity from the star is indicated via the yellow arrows. Magnetic fields of star and planet (indicated with thin black curves) define the strength of induction heating.*

**Cooling of planetary bodies**

For a solid surface, cooling of the planet is mostly limited to conductive heat flow $q$ at the surface through the lithosphere and crust, with the exception of direct transport of lava flows to the surface (in the case of volcanic activity, or, in the more extreme case, heat-pipe mechanism), which can lead to high local heat fluxes as seen on Io. The heat flux depends on the local thermal conductivity $k$ as well as the temperature gradient through the lithosphere and crust.

(7) $q = -k \frac{\partial T}{\partial r}$

On Earth, the heat flux varies between ~65 mW/m² (continental crust) to ~105 mW/m² (oceanic crust). For the moon, Apollo measurements showed a much lower heat flux in the range of 15-20 mW/m². In contrast, Io's heat flux has been measured to be about 2000 mW/m² due to the strong interior heating by tidal dissipation.

For the general efficiency of cooling of the interior (including secular cooling, i.e. the cooling initial accretional and gravitational heat sources, but also cooling of produced internal energy from tidal heating, radiogenic heating, or induction heating), smaller bodies should cool much faster than more massive planets, since the cooling occurs over the planetary surface, and the ratio of planet surface to planet volume (i.e. heat) decreases with increasing radius.

## 3.2. Interior as driver of planetary dynamics

Heat sources as described above are the main driver of planetary geodynamics, where thermally-induced buoyancy in core or mantle leads to the initiation of convective currents - in the core one of the necessary ingredients to drive a magnetic field. A planetary dynamo is typically only expected for cooling cores. For a strongly cooling core (typically assumed for the early evolution of low-mass rocky planets), the top of the liquid metal core will become denser than the underlying, warmer core, and convective currents are triggered to establish again a gravitationally stable field, i.e. lighter material lying on top of heavier material. A dynamo created by such a strong heat flux at the core-mantle boundary is termed a thermal dynamo.

On Earth, today the heat flux at the core-mantle boundary would not be sufficient anymore to maintain a magnetic field. However, cooling of the core over several Gyr led to the freezing of an inner, solid core. Since lighter elements in the core (such as various volatile elements) remain preferentially in the liquid upon solidification of iron and other metals, the layer above the solid inner core is enriched in lighter elements, hence the layer is lighter than the rest overlying liquid core, again triggering convection. A dynamo driven by the chemical variation of different core layers is accordingly referred to as chemical dynamo. Other types of chemically-driven dynamos exist, such as the iron-snow regime (especially relevant for low-pressure environments, e.g. for the core of Ganymede), and depend on the chemical composition and pressure-dependence of the adiabatic temperature profile in comparison to the melting temperature (Breuer et al., 2015).

We also expect convection in the solid, rocky mantle, which is then similarly driven by thermal or chemical density variations, but convection in the mantle occurs on much longer, geological timescales. Indications for the existence of mantle convection (i.e. that the mantle indeed behaves as a viscous fluid on geological timescales) go back to the observation of plate motion at the surface of Earth as well as seismic tomography indicating upwelling mantle plumes. Experimental rheological studies apply large stresses to upper mantle rocks or analogue materials and can be used to derive the viscosity of different materials. For rocks under higher pressure, theoretical studies can investigate first-order deformation mechanisms and derive rheological laws applicable to the deep interior of Earth, and potentially of super-Earths, as well.

Convection in a compressible mantle is typically described with the following so-called truncated anelastic liquid approximation, describing the conservation of mass, momentum and energy.

(8) $\nabla \cdot (\rho v) = 0$

(9) $-\nabla p + \nabla \cdot \sigma = Ra\, \rho\, g\, \alpha(T - T_{ref})e_r \qquad \sigma = \eta\left(\nabla v + \nabla v^T - \frac{2}{3}\nabla \cdot v\, I\right)$

(10) $\rho C_P \left(\frac{\partial T}{\partial t} + v \cdot \nabla T\right) = \nabla \cdot (k\nabla T) + Di\, \alpha\, g\, v_r(T + T_0) + \frac{Di}{Ra} \cdot 2\eta \left\|\nabla v + \nabla v^T - \frac{2}{3}\nabla \cdot v\right\|^2 + \rho H$

These equations depend on several thermodynamic parameters (such as density $\rho$, heat capacity $C_P$ and thermal expansion coefficient $\alpha$), transport properties (such as viscosity $\eta$ and thermal conductivity $k$), and variables of interest (temperature $T$, velocity vector $v$, and convective pressure $p$), as well as various heat sources $H$ as described above and geometric factors such as the radius union vector $e_r$ and union matrix $I$. There are well-established numerical routines to solve such a system of equations. However, parameters and variables are defined for very different orders of magnitude (with viscosities in the order of $10^{20}$ Pas and velocities in the order of $10^{-12}$ m/s. It is therefore an established approach to non-dimensionalize the conservation equations, as was done already in the equations (8)-(10) above, i.e. dividing each quantity by a reference value to yield non-dimensional values in the order of 1. With this approach, new quantities appear as left-overs of the non-dimensionalization process, that can give a first-order characterization of the overall strength of convection (as in the case of the Rayleigh number $Ra$) or of the compressibility of the mantle (expressed via the Dissipation number $Di$):

(11) $Ra = \frac{\rho_{ref} g_0 \alpha_{ref} \Delta T D}{\kappa_{ref} \eta_{ref}}, \quad Di = \frac{\alpha_{ref} g_0 D}{C_{P,ref}}$

where $g_0$ is the surface gravitational acceleration, $\Delta T$ a reference temperature contrast (such as initial CMB temperature minus surface temperature), $D$ is a measure of convection length (such as difference between planet and core radius), and $\kappa = k/(\rho\, C_P)$ is the thermal expansivity. Especially the Rayleigh number is of high importance to obtain first estimates on the strength of convection, where

(12) $v \approx Ra^{2\beta}$,

or in an extended definition (Valencia et al., 2007),

(13) $v \approx \frac{\kappa}{D} Ra^{2\beta}$,

as well as for the strength of the surface heat flow, expressed as Nusselt number, indicating the strength of convective heat flow over convective and conductive heat flow at the planet's surface,

(14) $Nu \approx Ra^{\beta}$.

The exponent $\beta$ has been derived in various laboratory, theoretical and numerical studies and lies approximately between 1/4 and 1/3 (depending on convection regime and geometry of the model domain).

The above defined dimensionless quantities (Rayleigh number, Dissipation number, and Nusselt number) are commonly used in fluid dynamics to characterize fluid properties and are used to derive scaling relationships describing the general dynamical behaviour of a convective medium. Since rocks behave similarly to fluids on geological time scales, these scaling laws allow to compare convection inside of a rocky mantle to small-scale laboratory experiments using fluids with different rheological properties.

An additional relationship has been observed between the Rayleigh number and the thickness of the thermal boundaries $\delta$ of a convecting layer, which are the layers forming between the actively convective region and the top or bottom boundary of the layer, and where thermal instabilities occur that lead to local overturns, driving convection in the layer,

(15) $\delta \approx D \cdot Ra^{-\beta}$.

Following Valencia et al. (2007), the total stress underneath the lithosphere of a planet can be approximated with

(16) $\tau \approx \eta v/D$,

which is related to the horizontal normal stress, that is needed for the lithosphere to break,

(17) $\sigma \approx \tau \delta v/\kappa$.

If the convecting mantle $D$ as well as convecting velocities $v$ increase with planet size, this would imply an increase in the likelihood of plate tectonics with planet size. However, other studies (summarized in Ballmer and Noack, 2021) have shown, that for increasing internal temperatures (as expected with increasing planetary mass), the likelihood of plate tectonics may decrease. On the other hand, such scaling laws as described above depend strongly on the thickness of the convecting layer, which is not necessarily the mantle thickness. For strongly increasing viscosities with pressure, the convecting layer of the mantle may actually decrease with increasing planet mass, leading to thicker lithospheres, lower mantle velocities and lower stresses, favouring a stagnant-lid scenario for super-Earth planets. Plate tectonics may therefore be more or less likely for increasing planetary mass, depending on several factors such as heat sources or pressure-dependence of the viscosity, and no linear trend can be observed.

The thickness of a stagnant upper layer (i.e. the lithosphere) is not only relevant for the prediction of the surface regime (plate tectonics vs stagnant lid), but also influences if melt can reach the surface though a stagnant lid or remains trapped at the base or inside the lithosphere. Depending on the assumed rheology in the interior, the predictions of melt volumes can therefore also be higher or lower for more massive planets. High-pressure rheological studies for various mantle compositions may therefore give a hint at what to expect for massive super-Earths - bare-rock, stagnant-lid surfaces without strong volcanic activity, or Earth-like worlds with diverse atmospheres fed by continuous plate-tectonics-driven volcanic outgassing.

In the end, observations of active volcanism on rocky exoplanets (i.e. by measuring traces of recent, instable volcanic compounds in the atmosphere or by observing bare-rock planets with strong interior heating) for planets of various planet masses may tell us in the future how the interior state as well as outgassing strength varies with planet mass, composition and orbital configuration (influencing e.g. tidal heating and induction heating of the planet).

### 3.3. Interior as source of crust and atmosphere

Such observations - especially of bare-rock planetary surfaces - can also give us additional insights on the interior of selected rocky exoplanets. Specifically, the composition of the crust may reflect at least to some part the composition and therefore mineralogy of the upper mantle of these bodies, where melt is formed by partial melting of rocks. On Earth, we distinguish rocks (mantle and crust) in four basic categories: ultramafic, mafic, intermediate, and felsic crust. An ultramafic composition is assigned to primitive mantle rocks, whereas mafic (basaltic) crust stands for a rock that is magnesium-rich and silicon-poor, i.e. composed of minerals that have a lower melting temperature in the mantle, therefore accumulating first in partial melt. Felsic crust describes a high rock with high feldspar and silicon contents, being therefore relatively magnesium-poor. Such a crust can contain large quantities of quartz ($SiO_2$) and is in general less dense than a mafic crust. Intermediate rocks contain a silica fraction between mafic and felsic crust. Prominent examples of the different rock types are peridotites (ultramafic mantle rocks), basalts (mafic crust, which is the dominant crustal rock of rocky planet surfaces in the solar system), as well as granites (felsic crust). Intermediate and felsic crusts are typically thought to be formed by re-melting of crust, driven by plate tectonics and subduction of crust and volatiles in the mantle.

The oceanic crust of Earth actually consists of basalts, and the continental crust is mostly made of felsic material. However, the same analogy cannot be transferred to our neighbor planets, as $SiO_2$-rich rocks have been found embedded in basaltic crust on both Venus and Mars, with no clear link to a continental crust, plate tectonics or recycling and remelting of crust. Similarly, our main classification of rocks (from ultramafic to felsic) may fail on planets with different initial compositions around stars that are for example by nature magnesium-rich or silicon-rich. In the light of new observational capabilities for exoplanet surface rocks, more experiments for various mantle compositions as well as pressure- and temperature conditions are needed to understand the link between mantle and crust for non-Earth-like compositions, as well as to constrain their spectral features.

In general, depending on the exact melt composition, structure, porosity and various weathering processes, crustal rocks can come in a rich variety of colors and major components. Care should therefore be taken when trying to characterize the planetary surface of a bare-rock exoplanet from spectral analysis, and especially when interpreting observations with respect to crust heritage and implications for surface recycling.

The same caution should be applied for the interpretation of atmospheric gases and their potential link to the interior of a planet. First, there is need to distinguish different types of atmospheres, varying mostly with planet mass and age. Second, all processes shaping a planetary atmosphere need to be considered, including internal and external sources as well as sinks of atmospheric gases.

We can broadly distinguish three types of atmospheres: The first type is a **primordial atmosphere**, that is formed by gravitational attraction from the planetary disk if the planet reaches a critical mass during the lifetime of the disk. This atmosphere should roughly resemble the stellar atmospheric composition (as is the case for the gas and ice giants in the solar system) with some variations with orbital distance and hence local temperature. These atmospheres are dominated by hydrogen and helium, but can also contain other gases including for example water, ammonia and methane (methane being the gas that gives Neptune and Uranus their blue color).

For young, rocky planets, and potentially also for sub-Neptune planets, the atmospheric composition is influenced by interactions with the interior, at least as long as the rocky mantle is molten at the surface and allows for volatile exchange with the atmosphere. This interaction may be a mixture of first-order chemical equilibrium, kinetic effects, saturation limits in the atmosphere, combined with solubility constraints for volatiles in the deep magma ocean. Due to a continuous exchange of volatiles between interior and atmosphere, especially during a potential solidification of the magma ocean leading to an enrichment of volatiles in the remaining magma ocean, a **primary outgassed atmosphere** forms, which can potentially also contain still large fraction of the primordial atmosphere - depending on the efficiency of atmospheric escape to space. It should be noted that almost none of the primordial hydrogen would be expected to be dissolved in the magma ocean due to the low solubility of hydrogen compared to other volatiles. Similarly, helium is incompatible with melts and solidified rocks and can only be a trace element in a planetary mantle and trace gas in planetary atmospheres that lost their primordial atmosphere.

For low-mass rocky planets, atmospheric escape may lead to the removal of not only the primordial, but also of the primary outgassed atmosphere after magma ocean solidification - depending on the surface gravity, stellar activity, and atmospheric composition. While Mars and Earth do not show any survival of their earliest atmospheres, and Mercury as well as the Moon do not contain any considerable atmosphere at all, anymore, for Venus the debate is ongoing. Was its current atmosphere mostly or entirely built by **secondary outgassing**, i.e. volcanic activity, or is some of it a relic of the primary outgassed atmosphere? Better atmospheric measurement of several trace gases in the next decade with the upcoming new Venus missions may shed a better light on the origin of Venus' atmosphere - and depending on the outcome, may finally help to answer the question if Venus was ever in a habitable phase or was always a hellish world, with $CO_2$ levels always beyond the runaway greenhouse point.

The different stages of atmospheric evolution are depicted with the example of our Earth in Fig. 4, where the early stages of a possible primordial atmosphere as well as the primary outgassed atmosphere on top of a magma ocean spanned in reality a much shorter time frame than depicted in the figure (for better visibility). The long-term atmospheric evolution was shaped first by volcanic outgassing (secondary outgassed atmosphere), followed in the case of Earth by a fourth atmospheric type - the **tertiary atmosphere** shaped by photosynthetic life leading to the great oxidation event.

From an observational point-of-view, however, it is not an easy task to differentiate between primordial or outgassed atmospheres, or any hybrid atmosphere stage in-between. One of the key elements in primordial atmospheres, however, is helium. In principle, the detection or non-detection of He in an exoplanet atmosphere would indicate if it is of primordial origin or not. However, detection of helium is highly challenging and currently restricted to exoplanets with an escaping atmosphere around stars with a specific radiative signature to make the helium visible. For now, theory therefore needs to predict how atmospheric types should differ for various planet masses and orbital configurations, including predictions on outgassing strengths from the interior of the planets (either during or after the magma ocean stage).

For both the primary and secondary outgassed atmospheres, the interior dynamics, energy budget, and chemistry indeed play major roles in shaping the final atmospheric composition as well as atmospheric pressure. However, several external factors further influence the long-term evolution of the atmosphere, including the stellar insolation (influencing the temperature and extend of the atmosphere, and therefore thermally-driven escape processes), stellar flares and CMEs stripping part of the atmosphere (decreasing strongly with distance), as well as the twofold effect of impacts - ranging from late veneer addition of volatiles to an atmosphere (potentially strongly changing the composition of the atmosphere) to the destructive potential of a larger impactor, stripping part of the atmosphere from the planet rather than increasing the atmospheric volatile content.

The main principles of atmospheric losses (including the importance of a magnetic field for some of the non-thermal loss processes) are a complex topic deserving their own review chapter. Here it should suffice to say that one of the main components for the survivability of an atmosphere is the composition of the atmosphere, where atmospheres with higher mean molecular weights (such as Venus' $CO_2$-dominated atmosphere) are more resistant against atmospheric escape, whereas hot, extended atmospheres (such as $H_2$-dominated atmospheres) are prone to atmospheric escape independent of the existence of a magnetic field due to thermal escape processes.

The main factor influencing, however, the dominant species in the atmosphere (as long as it is not of primordial origin), is volcanic outgassing from the interior - depending not only on the volatile composition of the mantle, but also the redox state of the mantle rocks and melts from which volatiles are degassed at the planetary surface. The redox state strongly influences the gas speciation and solubility of volatiles in the mantle or melt. In addition, outgassing from surface magma is also limited by the atmosphere itself, more precisely by the partial pressures of individual gases in the atmosphere. For dense atmospheres, water and sulfur species may remain dissolved in the magma and not further contribute to the atmosphere, whereas other species such as $CO_2$ or $H_2$ degas easily and can build up very dense atmospheres.

Which gases ultimately enrich an atmosphere is therefore not directly linked to the redox state of the melt (or at least not only), and measurements of atmospheric compositions therefore do not allow for a linear link to interior chemistry - though endmember atmospheres (e.g. very reducing or very oxidizing atmospheres) may shed a first light on the mantle chemistry and composition. It should also be mentioned that the atmosphere is of course also prone to changes by several additional surface and atmospheric processes, including atmosphere losses (e.g. loss of $H_2$ oxidizing the remaining atmosphere), various chemical pathways in the atmosphere, equilibrium vs. disequilibrium considerations, condensation and formation of clouds or water oceans, weathering and chemical reactions at the surface, recycling of surface reservoirs, global feedback cycles such as the carbon-silicate cycle, and last but not least, potentially, due to the influence of life.

**Conclusion**

The field of exoplanetary research has seen a tremendous change from first-order characterization of planets based on mass and radius measurements to the investigation of the complex and strongly interlinked evolutionary pathways of planetary interiors, surfaces and atmospheres by studying planets in the context of their environment (especially with respect to composition), interior dynamics (from accretion to long-term evolution of planets) and feedback links between the interior and the atmosphere. Future, more detailed atmospheric characterization surveys of planets in multiplanetary systems, as well as planets over large parameter spaces including stellar diversity, variable ages and different orbital configurations, will allow us to test our theoretical predictions and better understand the place of our own solar system planets within the exoplanetary context.

**Acknowledgements**

Funded by the European Union (ERC, DIVERSE, 101087755). Views and opinions expressed are however those of the author(s) only and do not necessarily reflect those of the European Union or the European Research Council Executive Agency. Neither the European Union nor the granting authority can be held responsible for them. C.D. acknowledges support from the Swiss National Science Foundation under grant TMSGI2_211313.